\documentclass[twocolumn,prl,showpacs]{revtex4}
\usepackage{amsmath}
\usepackage{graphicx}

\newcommand{\eq}{\begin{equation}}
\newcommand{\fine}{\end{equation}}

\begin{document}

\title{Towards Quantum Experiments with Human Eyes Detectors Based on
Cloning via Stimulated Emission ?}
\author{Francesco De Martini}
\affiliation{Dipartimento di Fisica, Sapienza Universit\'{a} di Roma, Roma, 00185 Italy}
\affiliation{Accademia Nazionale dei Lincei, via della Lungara 10, I-00165 Roma, Italy}

\begin{abstract}
We believe that a recent theoretical work published in Physical Review
Letters (103, 113601, 2009) by Sekatsky, Brunner, Branciard, Gisin, Simon,
albeit appealing at fist sight, is highly questionable. Furthermore, the
criticism raised by these Authors against a real experiment on Micro - Macro
entanglement recently published in Physical Review Letters (100, 253601,
2008) is found misleading and misses its target.
\end{abstract}

\maketitle

\subsection{\protect\bigskip}

We believe that the work by P.Sekatsky, N.Brunner, C. Branciard, N. Gisin
and C.Simon is highly questionable \cite{Bran 09}. The first seed of
perplexity is elicited by the title of the paper, the same as the one of the
present article (apart from the question mark). For the eye of a human
observer, as well as any other human sensory organ, just cannot be\ adopted
as a valid \textquotedblright measurement apparatus\textquotedblright\
within any experiment involving a quantum mechanical process. As stated many
times by Niels Bohr, this apparatus must be a \textquotedblright
classical\textquotedblright\ one, i.e. whose behavior\ follows reliably the
well established, deterministic laws of classical physics\cite{Bohr}. This
is vividly expressed by the prose of J.A.Wheeler \cite{Wee}:
\textquotedblright A phenomenon is not yet a phenomenon until is a
registered phenomenon i.e. brought to a close by an irreversible act of
amplification\textquotedblright . The measurement apparatus must be composed
by\ three devices: (a) A detector, e.g. a photocathode or a grain of silver
bromide etc, that realizes the reduction of the quantum wavefunction, (b) A\
classical amplifier (c) A registering unit, or a memory that records the
outcome of the measurement, i.e. a \textit{real number}. Indeed the last
device is a utterly necessary item in order \textquotedblright to bring to a
close\textquotedblright\ the measurement and then to establish
\textquotedblright the phenomenon\textquotedblright . We remind here that
the \textquotedblright memory erasure\textquotedblright\ (or the
\textquotedblright register re-setting\textquotedblright ) was the necessary
conceptual step taken by Charles Bennett in order to resolve the famous
\textquotedblright Maxwell demon Paradox\textquotedblright\ \cite{Benn} In
short, a measurement\ is far more than a mere perception, i.e. a solipsistic
process. Any measurement outcome must be available to the \textquotedblright
observer\textquotedblright\ as well as to an independent scientific
community by which it can be promoted to the level of \ \textquotedblright
datum\textquotedblright\ ready to be adopted by a scientific theory. Quite
obviously, the eye detection does not comply with these requirements for
several reasons. First, the \textquotedblright recording
device\textquotedblright\ is not in front of but rather literally \textit{in
the head} of the observer. Then, to say the least, the measurement outcomes
cannot be independently addressed by other observers. Second, we couldn't
imagine which "mental pointer" or mental counter or mental scaling algorithm
could be adopted in order to transform, in a reliable and reproducible way,
the level of the synaptic electric field\ into one amongst a set of \ 
\textit{orthogonal} \textit{outcomes}\ expressing the detected light
intensity, i.e. the only \textquotedblright observable\textquotedblright\
accessible to the eye. In facts the "pointer" sitting somewhere within the
brain of the observer should be able to single out a definite orthogonal
outcome (a), say: a = 3, rather than: a = 2, or a = 4. Third, while the role
of the retina and the Na - ion excitation dynamics of the optical nerve may
be taken as rather well understood, the complex synaptic trasmission to the
\textquotedblright reentrant\textquotedblright\ talamo - cortical system of
the brain \cite{Edel} and the consequent establisment of the various levels
of memory \cite{Kandel} are\ obscure and are today the subject of frontier
research within the domain of the advanced neurosciences. For the present
purpose we only know for sure some\ phenomenological properties of the
brain. For instance,\ that the optical nerve amplification is highly
nonlinear and that the overall visual efficiency is easily saturated.
Furthermore, some drugs\ as alchool (contained in wine) and betacarotene
(contained in carrots or tomatoes) not to speak of other more dangerous
drugs can have large and opposite effects on the parameters of visual speed,
nonlinearity and efficiency. Indeed too many scarcely controllable things
can happen simultaneously within our head in any moment: perceptions,
emotions, desires, phantasies, rational and irrational thoughts: all that is
part of the rich realm of "consciousness". Then, while we largely sympatize
with the open mindness revealed by the work done in the domain of \
measurement related consciousness by scientists as Von Neumann\cite{Neumann}%
, Pauli\cite{Pauli}, Wigner \cite{Wigner}, Stapp \cite{Stapp} and others, we
can't restrain from manifesting our opposition when dealing with serious
real experiments and theories. In summary, we believe that today, and
foreseeably for a very long time in the future, the human eye detection
should be thought of as a largely useless \textquotedblright epistemic
mess\textquotedblright\ when related to the argument of quantum measurement.

\begin{figure}[t]
\includegraphics[width=0.50\textwidth]{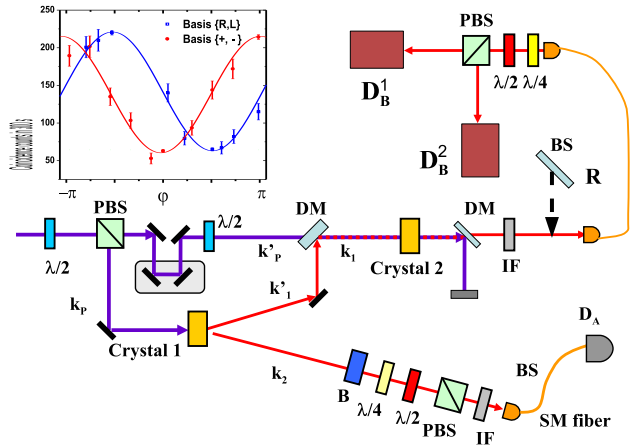}
\caption{Layout of the Micro-Macro experiment published in Phys.Rev.\ Lett.
100, 253601 (2008). \ At the output of the QI-OPA\ Amplifier an optional
beam-splitter with reflectivity R simulates the photon - loss in the
detection process. \ In order to determine the visibilities $V_{2}$ and $%
V_{3}$ , the couples of experimental points corresponding to the maxima and
minima of the fringing patterns obtained in correspondence with two
different measurement bases have been determined by higher statistics and
exhibit a smaller error flag. }
\label{fig:concurrence}
\end{figure}

The bravery of P.Sekatsky et al. with their eye detection proposal is
further revealed by the \textit{gedanken experiment} considered in\cite{Bran
09} and dealing with a nonlocality test made on a micro-macro photon system
by the violation of a Bell inequality. We may describe the proposed
experiment on the basis of their Figure 3, as follows. A standard laser
system generates by Spontaneous Parametric Down Conversion (SPDC) a
polarization \textit{singlet} couple of photons that are sent to two
spacelike distant stations, Alice and Bob. The single photon received by
Alice is measaured by a standard Optical Stern-Gerlach apparatus $(OSG_{A})$
consisting of a couple of single-photon detectors coupled to the two output
modes of a Polarizing Beam Splitter $PBS_{A}$. As usual in these
experiments, the $OSG_{A}$ is "rotated" of an angle $\Phi _{A}$ (All
generalized "rotations" considered in the present letter could possibly
imply changes of the state of photon polarization, e.g. from "linear" to
"circular" etc.) \ Likewise, the single photon send towards Bob, is
amplified by some unspecified device. The $N$ \ photons generated by the
amplifier are then measured by another Optical Stern-Gerlach system $%
(OSG_{B})$ consisting of a $PBS_{B}$ "rotated" by an angle $\Phi _{B}$. \
Needless to say, the $(OSG_{B})$ is completed by two naked human eyes,
accurately drawn in the Figures 1 and 3 of \cite{Bran 09}, each one staring
in one of the output modes of $\ $the$\ PBS_{B}$. This proposed \textit{%
gedanken experiment} reproduces almost exactly the \textit{real experiment}
previously carried out by \cite{DeMa08} where the amplifier is a Quantum
Injected Optical Parametric Amplifier (QI-OPA) generating $N\simeq 10^{5}$
output photons \cite{DeMa98} and the human eyes are replaced, perhaps more
reasonably, by two detection apparata $D_{B}^{1}$, $D_{B}^{2}$ each
involving a linear photomultiplier $.$ The experimental layout\ of our%
\textit{\ real} experiment is shown in Figure 1, above.

The gedanken experiment planned by Sebatsky \textit{et al}. raises further
obvious questions. For instance, the two naked eyes, in order to be able to
measure different signals, cannot belong to the same person because of the
physiological \textit{fusion} of the related perceptions due to the
inextricable and incontrollable interconnections between the optical nerves
reaching the left and right sectors of the same brain. The eyes must then
belong to two different persons (two students ?). At that point, since the
Bell inequality experiments with many photons cannot imply simple yes/no
responses but require the registering of the actual level of the\ synaptic
signals, we are again confronted with slippery unanswerable questions about
mental pointers, mental signal processing, saturation, linearity, neural
connections, betacarotene and alchool, mutual calibration and stability,
consciousness etc.

At last, let's stop pondering on the bizarre naked eye detection idea and do
consider the detailed micro-macro Bell inequality theory also reported in 
\cite{Bran 09}.\ There a measurement \textit{loophole} is devised in
physical situations implying the calculation of the joint correlation\
parameters between apparata $(OSG_{A})$ and $(OSG_{B})\ $tuned on different
measurement bases, i.e. when the relative angular settings\ of the
corresponding measurement apparata differ from zero: $\Delta \Phi \equiv
\left\vert \Phi _{A}-\Phi _{B}\right\vert $ $\neq 0$. Indeed, this is a
typical situation realized in all Bell inequality experiments.\ We don't
disagree on several results of the theoretical analysis by \cite{Bran 09}
but we also want to stress that these ones are quite incorrectly applied to
the \textit{real} experiment reported in \cite{DeMa08}. In other words, the
criticism \textit{\ }to our work by Sebatsky \textit{et al}, presumably the
true motivation of work \cite{Bran 09}, is misleading as it misses
completely the point. For the following reasons:

(A) The work \cite{DeMa08}\textit{\ is not} a Bell inequality experiment and
then no correlations between \textit{different} measurement bases are
measured or calculated within the same experiment.. The work \cite{DeMa08}
merely consists of two totally \textit{independent and uncorrelated} \
experiments aimed at the evaluation of two \textit{different and uncorrelated%
} quantities, i.e. the "visibilities" $V_{2}$ and $V_{3}$ of the two \ 
\textit{different\ and uncorrelated} \ fringing patterns shown in Figure 1,
above. (The other "visibility" was found: $V_{1}\simeq 0$). These patterns,
drawn as function of $\Phi _{B}$, represent the jointly correlated detection
probabilities when a fixed measurement basis of $(OSG_{A})$ is chosen to be
either $\left\{ R,L\right\} $ or $\left\{ +,-\right\} $, respectively.
Consider for instance the measurement of $V_{2}$, i.e. the visibility\ of
the fringe pattern determined by the fixed basis $\left\{ R,L\right\} \ $set
at the Alice's site. As it is well known $V_{2}$ is determined by only two
points, the \textit{maximum} and the \textit{minimum} of the pattern, i.e.
exactly the points corresponding to the conditions: $\Phi _{A}$ $=$ $\Phi
_{B}$, or:\ $\Delta \Phi =0.$In other words, the two data used to evaluate $%
V_{2}$ are obtained by measuring the joint detection probabilities in the
conditions in which the micro-qubit at Alice's site and the macro-qubit at
Bob's site are mutually parallel or anti - parallel spin vectors i.e. both
belonging to the \textit{same} $\left\{ R,L\right\} $ basis on the
corresponding, equally oriented Poincar\'{e} spheres. \ \ The same
condition: $\Phi _{A}$ $=$ $\Phi _{B}$, or:\ $\Delta \Phi =0$ \ is realized
within the measurement of $V_{3}$ where again the common measurement basis $%
\left\{ +,-\right\} $ is realized for both the Alice's and Bob's apparata.
Then, because of the common condition: $\Delta \Phi =0$ affecting both
measurements of $V_{2}$ and $V_{3}$, the "loophole" devised by Sekatsky 
\textit{et al}. is not applicable to our experiment.

(B) Symmetry considerations based on the \textit{rotational invariance} of
the overall micro-macro \textit{singlet} photon pair expressed by Equation 1
in \cite{DeMa08}, and of the \textit{phase-covariant} and \textit{%
information preserving} properties of the of the adopted QI-OPA,\ lead to
conclude that the two $V_{2}$ and $V_{3}$ experiments are really identical,
in the sense that the micro and macro states adopted in both cases, albeit
formally different, are in fact obtained by relabelling for different
polarizations the Fock state components of these micro and macro-states. In
facts, the experimental outcomes $V_{2}$,$V_{3}$ of the two corresponding
experiments have been found equal by \cite{DeMa08}, within the statistical
errors.

\begin{figure}[t]
\includegraphics[width=0.40\textwidth]{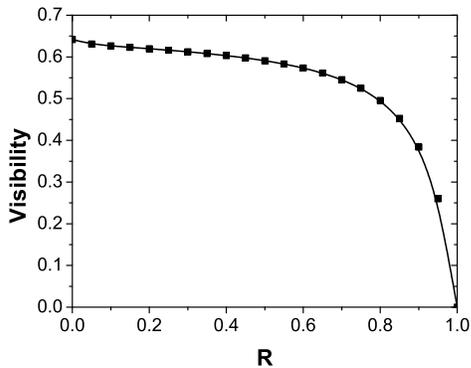}
\caption{Complete computer simulation of the experiment [12] showing the
decrease of \ the visibility $V_{2}$ \ due to the reduction of micro - macro
entanglement for increasing R, i.e. the amount of photon loss. . }
\label{fig:concurrence}
\end{figure}

(C) As presumed by Sekatsky \textit{et al}, photon losses are indeed present
in the multi-photon (Bob) side of experiment \cite{DeMa08}, mostly due to
the reduced quantum efficiency $QE<1$ of the photomultipliers. In any case\
the effect of \ losses is a "\textit{local"} one and may be modelled, as
shown above in Figure 1, by a Beam Splitter (BS) with a transmission $%
T\equiv (1-R)$ placed right at the output of the QI-OPA apparatus. The
result of a complete computer simulation of the experiment \cite{DeMa08} by
adopting the real experimental parameters and by assuming the fixed
measurement basis $\left\{ R,L\right\} $, is shown in Figure 2. There the
"visibility" $V_{2}$, reported as function of $R$. is found to be a \textit{%
decreasing} function of of \ the amount of \ photon losses. This result is
expected since, being the micro-macro entanglement distributed between all
photons emitted by QI-OPA, any photon loss entails a reduction of the amount
of entanglement detected on the remaining photons. Furthermore, this
behavior \ agrees with a nice "\textit{entanglement criterion}" expressed in
a paper by Eisenberg \textit{et al} \cite{Eise} that can be expressed as
follows: "any \textit{local} transformation cannot enhance the level of
entanglement". A photon loss is indeed a \textit{local} transformation, by
definition. The work by Eisenberg et al. \cite{Eise} also dealt with
experimental multiphoton etanglement detection with $QE<1$\cite{simon}.$.$

In spite of the entanglement reduction due to the\ measurement losses, the "%
\textit{visibility inequality}" $\left\vert V_{1}+V_{2}+V_{3}\right\vert
\leq 1$ \ was violated in the experiment \cite{DeMa08}. This \textit{a
fortiori} demonstrates the \textit{nonseparability} of our Micro - Macro
system.

In summary, all previous considerations fully support our claim asserting
that the work \cite{DeMa08}, taken together with previous works by our
Laboratory \cite{DeMa98}\cite{DeMa00} indeed consists of the first \textit{%
exact} realization of the \textit{Macroscopic Quantum Superposition}, i.e.
complying \textit{exactly} with the original definition given by Schr\"{o}%
dinger in 1935 \cite{Sch39}. The value of this discovery is further enhanced
by the large resilience to decoherence shown by our system, which involves
as many as $N\simeq 10^{5}$ particles \cite{DeMa09}. The robustness against
any kind of noise makes our system apt to the investigation on several so
far inaccessible fundamental issues of quantum mechanics close to the
elusive "\textit{quantum - classical boundary}".

We conclude by stressing our deep appreciation for the continuous interest
in our work by P.Sekatsky, N.Brunner, C.Branciard, N. Gisin and C. Simon.


\bigskip

\begin{thebibliography}{10}

\bibitem{Bran 09} P. Sekatsky et al., Phys. Rev. Lett. \textbf{103}, 113601 (2009).

\bibitem{Bohr} N. Bohr, Atomic Physics and Human Knowledge (Ox Bow Press, Woodbridge, Conn. 1963).

\bibitem{Wee} J.A. Wheeler, Law without Law in: Quantum Theory and Measurement, J.A. Wheeler and W.H. Zurek eds. (Princeton University Press, 1983).

\bibitem{Benn} C.H.Bennett, Int. J. Theor .Phys. \textbf{21}, 905 (1982); Sci. Am. \textbf{257}, 108 (1987).

\bibitem{Edel} G. M. Edelman, Mindful Brain: Cortical Organization and the Group - selective theory of higher Brains (MIT Press, Cambridge 1978).

\bibitem{Kandel} E. R. Kandel, In Search of Memory (Norton, New York 2006).

\bibitem{Neumann} J.Von Neumann, \textit{Matematischen
Grundlangen der Quantenmechanik} (Springer, Berlin 1932)

\bibitem{Pauli}H. Atmanspacher, H.Primas, \textit{%
Recasting Reality} (Springer, Berlin 2009).

\bibitem{Wigner} E.P. Wigner, Symmetries and Reflections (University of Indiana, Bloomington 1967)

\bibitem{Stapp} H.P: Stapp, Mind, Matter and Quantum Mechanics (Springer, Berlin 1993).
\bibitem{DeMa98} F. De Martini, Phys. Rev. Lett. \textbf{81}, 2842 (1998); Phys. Lett. A \textbf{250}, 15 (1998).


\bibitem{DeMa08} F. De Martini, F.Sciarrino, and C. Vitelli, Phys. Rev. Lett. 
\textbf{100}, 253601 (2008).

\bibitem{Eise} H. Eisenberg, G. Khoury, A. Durkin, C. Simon, and D. Bouwmeester, Phys. Rev. Lett. \textbf{93}, 193901 (2004).

\bibitem{DeMa00} F. De Martini, F. Sciarrino, and V. Secondi, Phys. Rev.
Lett. \textbf{95}, 240401 (2005); F. De Martini, and F. Sciarrino, Progr. Quantum Electr.
\textbf{29}, 165 (2005); F. De Martini, and F. Sciarrino, Journal of Physics A: Math. Theor. \textbf{40},
2977 (2007).


\bibitem{Sch39} E. Schr\"{o}dinger, \textit{%
Naturwissenshaften} \textbf{23}, 807 (1935).

\bibitem{DeMa09} F. De Martini, F. Sciarrino and N. Spagnolo, Phys. Rev. Lett. \textbf{103}, 100501 (2009).  

\bibitem{simon} Since Christoph Simon co-authored
both papers \cite{Eise} and \cite{Bran 09} he may perhaps explain why \ the "%
\textit{loophole}" problems should be applicable to work \cite{DeMa08} and not to 
\cite{Eise} and why the "\textit{visibility inequality}" and the "\textit{%
entanglement criterion}" should be applicable to work \cite{Eise} and not to \cite{DeMa08}%.

\end{thebibliography}
\end{document}